\documentclass[manuscript, sigconf]{acmart}
\usepackage{makecell}
\usepackage{multirow}
\usepackage{xcolor}
\AtBeginDocument{%
  }

\copyrightyear{2025}
\acmYear{2025}
\setcopyright{cc}
\setcctype{by}
\acmConference[ICER 2025 Vol. 1]{ACM Conference on International Computing Education Research V.1}{August 3--6, 2025}{Charlottesville, VA, USA}
\acmBooktitle{ACM Conference on International Computing Education Research V.1 (ICER 2025 Vol. 1), August 3--6, 2025, Charlottesville, VA, USA}
\acmDOI{10.1145/3702652.3744208}
\acmISBN{979-8-4007-1340-8/2025/08}

\begin{document}

\title[Validation of the Critical Reflection and Agency in Computing Index]{Validation of the Critical Reflection and Agency in Computing Index: Do Computing Ethics Courses Make a Difference?}

\author{Aadarsh Padiyath}
\email{aadarsh@umich.edu}
\orcid{0000-0002-4898-3566}
\affiliation{%
  \institution{University of Michigan}
  \city{Ann Arbor}
  \state{Michigan}
  \country{USA}
}

\author{Casey Fiesler}
\email{casey.fiesler@colorado.edu}
\orcid{0000-0002-8743-4201}
\affiliation{%
  \institution{University of Colorado}
  \city{Boulder}
  \state{Colorado}
  \country{USA}
}

\author{Mark Guzdial}
\email{mjguz@umich.edu}
\orcid{0000-0003-4427-9763}
\affiliation{%
  \institution{University of Michigan}
  \city{Ann Arbor}
  \state{Michigan}
  \country{USA}
}

\author{Barbara Ericson}
\email{barbarer@umich.edu}
\orcid{0000-0001-6881-8341}
\affiliation{%
  \institution{University of Michigan}
  \city{Ann Arbor}
  \state{Michigan}
  \country{USA}
}

\begin{abstract}
Computing ethics education aims to develop students' critical reflection and agency. We need validated ways to measure whether our efforts succeed. Through two survey administrations (N=474, N=464) with computing students and professionals, we provide evidence for the validity of the \textit{Critical Reflection and Agency in Computing Index}. Our psychometric analyses demonstrate distinct dimensions of ethical development and show strong reliability and construct validity. Participants who completed computing ethics courses showed higher scores in some dimensions of ethical reflection and agency, but they also exhibited stronger techno-solutionist beliefs, highlighting a challenge in current pedagogy. This validated instrument enables systematic measurement of how computing students develop critical consciousness, allowing educators to better understand how to prepare computing professionals to tackle ethical challenges in their work.

\end{abstract}




\maketitle

\section{Introduction}

Computing and software systems are ``woven into the fabric of everyday life'' \cite{weiser1999computer} from employment \cite{o2017weapons}, to justice \cite{benjamin2019race} to healthcare \cite{leveson1993investigation}. Thus, software developers have a growing responsibility to understand the ethical and sociotechnical implications of their work \cite{vakil2018ethics, ko2020time, yadav2022toward}. This is especially apparent within what critics call the \textit{``techlash''} \cite{su2021critical, sarder2022entering, ryoo2024show}: the growing public frustration with tech companies and their products. From backlash regarding algorithmic manipulation \cite{hunter2016facebook} and deceptive ``dark'' patterns \cite{gray2018dark} to widespread data misuse \cite{o2017weapons, zuboff2019age}, these controversies expose both the immense power of technology and the societal responsibilities of those who shape it.

Our profession has long recognized the importance of ethical awareness and sociotechnical understanding \cite{martin1996implementing}, even as the myth of computing as an apolitical, purely technical field has persisted in broader contexts \cite{vakil2018ethics, malazita2019infrastructures, benjamin2019race, ko2020time, ryoo2024show}. Encouragingly, the past decade has seen a dramatic shift, with educators, policy-makers, and the public increasingly recognizing the ethical dimensions of developing technology \cite{o2017weapons, noble2018algorithms, eubanks2018automating, gray2019ghost, zuboff2019age, benjamin2019race, brown2024teaching, padiyath2024realist}. Building on this, our community has sought to establish clear guidelines about ethical standards of our profession
and in computing education
\color{black}
through efforts such as the ACM Software Engineering Code of Ethics \cite{gotterbarn1997software}, the ABET accreditation criteria \cite{ABET_2024}, the ACM/IEEE/AAAI-endorsed CS2023 curricula \cite{SEP}, and dedicated computing ethics courses \cite{fiesler2020we}.

This growing interest in computing ethics education has sparked numerous interventions \cite{brown2024teaching} and theories \cite{morales2023conceptualizing} -- yet we still have not answered a fundamental question: 
Is it working? 
\color{black}
Prior research has qualitatively documented the challenges of engaging students in computing ethics \cite{padiyath2024realist}: Some students feel ethics material is too abstract, assuming it is disconnected from actual practice \cite{kirdani2022house}; Others struggle to understand how to navigate ethical issues when dealing with power dynamics in the workplace \cite{castro2023piloting}. While some students find ethics training valuable, others prioritize purely technical training \cite{padiyath2024realist}. Quantitatively assessing the impact of ethics courses has proven difficult due to a reliance on intervention-specific surveys which often lack theoretical grounding and psychometric evidence of reliability and validity \cite{brown2024teaching, padiyath2024realist}.

The \textit{Critical Reflection and Agency in Computing Index} \cite{padiyath2025development} shows promise in addressing this gap. 
Grounded in \textit{Critically Conscious Computing} theory \cite{padiyath2025development} (itself inspired by Freire's concept of \textit{conscientização} \cite{freire2020pedagogy, watts2011critical, diemer2017development}), it posits that 
\color{black}
students who are aware of socio-technical issues (\textit{critical reflection}) and capable of acting on them (\textit{critical agency}) will maintain a critically conscious computing practice (\textit{critical action}). 
The index was developed through rigorous psychometric methods \cite{boateng2018best} in order to provide a theoretically-grounded framework for measuring both students' current attitudes and potential development over time. 
\color{black}
Although its initial content validity results were promising \cite{padiyath2025development}, an argument for the psychometric validity of its underlying structure
(statistically demonstrating that the index truly captures the proposed latent constructs)
\color{black}
is necessary before widespread adoption. 
Therefore, our study is guided by the following research questions:

\textbf{RQ1}: What is the underlying factor structure of the Critical Reflection and Agency in Computing Index?

\textbf{RQ2}: Does the index demonstrate adequate reliability, construct validity, and criterion validity for use in computing education contexts?

\textbf{RQ3}: Do students who receive computing ethics and/or professionalism training demonstrate stronger critical reflection and agency compared to those who don't?

Guided by best practices for index development and validation \cite{boateng2018best}, we conducted a validation study spanning two survey administrations
($N_1$=474, $N_2$=464) with computer science students and professionals through the data collection platform Prolific
\color{black}
to examine the index's factor structure, internal consistency, construct validity, and criterion validity. We
also
\color{black}
explored relationships between scores and respondents'
prior
\color{black}
exposure to ethics coursework to gauge these courses' potential impact on respondents' 
current
\color{black}
ethical orientations. Our main contributions are:

\begin{enumerate}
    \item Evidence for the validity of the Critical Reflection and Agency in Computing Index, which provides researchers a reliable tool to systematically measure computing students' ethical development; and

    \item Analysis that finds computing ethics courses are associated with higher critical reflection and agency along some dimensions, suggesting our current approaches are cultivating essential skills.
\end{enumerate}

By providing evidence for the validity of this index, we equip educators with a theoretically grounded measure to identify patterns in students' ethical development, inform the design and implementations of ethics interventions, track attitudes over time, compare approaches across institutions, and build a shared understanding of best practices for computing ethics pedagogy.

\section{Background}

Ethics and issues around the social impacts of technology have long been considered essential components of computing education \cite{martin1996implementing}. Professional organizations and accreditors recognize this importance through ethical codes and standards \cite{ABET_2024, gotterbarn1997software}. However, computing educators face significant challenges when incorporating ethics into their teaching \cite{smith2023incorporating}. This challenge stems from what educators often prioritize -- technical algorithms over their societal implications -- leading to what \citeauthor{cech2014culture} describes as a ``culture of disengagement'' \cite{cech2014culture} where students perceive ethics as disconnected from technical practice. Without clear frameworks and measures, instructors may struggle to evaluate whether their pedagogical approaches effectively bridge this disconnect and develop students' ethical agency.

The theory of \textit{Critically Conscious Computing} \cite{padiyath2025development} provides a framework for understanding how computing students might develop an ethical practice. Drawing from Paulo Freire's concept of critical consciousness \cite{freire2020pedagogy}, this approach guides educators to cultivate both critical reflection (helping students recognize systemic issues in technology) and critical agency (building students' confidence to address these issues), which together lead to ethical action in professional contexts \cite{padiyath2025development}. Developing these skills together can offer instructors a structured approach to computing ethics education that goes beyond simply describing ethical dilemmas and Western ethical frameworks \cite{barkhuff2025towards}.

Despite educators' focus on fostering critical reflection and agency, 
systematic literature reviews of computing ethics pedagogy by \citeauthor{brown2024teaching} and \citeauthor{padiyath2024realist} both highlight that there has been limited use of psychometric 
\color{black}
surveys to measure whether our interventions actually achieve their intended outcomes \cite{brown2024teaching, padiyath2024realist}. While various assessment tools exist -- from general critical consciousness \cite{diemer2017development, diemer2022development} and ethical orientation measures \cite{canney2016validity} to computing-specific surveys \cite{horton2022embedding} -- they often lack either theoretical grounding, computing-specific contexts, and/or evidence of reliability and validity. This makes it difficult for instructors to evaluate, compare, and improve their teaching strategies for computing ethics \cite{brown2024teaching, padiyath2025development}.

The Critical Reflection and Agency in Computing Index \cite{padiyath2025development} addresses this need by providing a theoretically-grounded tool designed specifically for computing education. While the index's initial development and content validity followed psychometric best practices \cite{boateng2018best}, establishing its psychometric validity for broad use requires a rigorous empirical study. Our research provides evidence for validity while investigating a fundamental question for computing ethics education: Do our ethics courses effectively teach the critical reflection and agency they aim to develop? Our validation enables educators to better understand student ethical development and improve pedagogy across our discipline.

\section{Methods}

The validation of the Critical Reflection and Agency in computing scale closely followed best practices for validating psychometric scales collected by \citet{boateng2018best}. After item development, the authors describe two main steps to gather evidence for validity and reliability: (1) Scale Development, involving both administering the survey and extracting factors; and (2) Scale Evaluation, involving tests of dimensionality, reliability, and validity \cite{boateng2018best}.

\subsection{Scale Development}

\subsubsection{Sampling and Survey Administration}

We administered our survey through Prolific, a crowd-sourcing platform with a large population of participants with programming knowledge \cite{tahaei2022recruiting}. Our target population criteria included: fluency in English, United States residence, and education in computer science. The survey was advertised as a 5-minute task with \$1.20 compensation (\$14.40/hour rate, rated ``Great!'' by Prolific's standards).

To minimize measurement errors, we included three attention-check questions throughout the survey with instructions to choose a specific option. Using Google Forms, we structured the survey into three sections corresponding to our hypothesized constructs, with items presented in multiple-choice grid format using a six-point Likert scale (Strongly Disagree (1/6) to Strongly Agree (6/6)). Following recommendations of sampling at least 10 participants per scale item~\cite{boateng2018best}, we requested an initial sample of 500 participants as larger samples typically yield lower measurement errors, more stable factor loadings, and more generalizable results \cite{boateng2018best}. We administered a refined version of the scale to a second sample of 500 participants, following \citeauthor{boateng2018best}'s recommendations for establishing scale reliability~\cite{boateng2018best}. The second administration was conducted approximately three weeks after the initial survey and excluded items that showed poor loading or cross-loading based on exploratory factor analysis from the first administration.

\subsubsection{Extraction of Factors}

We conducted exploratory factor analysis (EFA) with
StataSE 18 to identify which survey items naturally clustered together in participants' response patterns, suggesting potential underlying factors of critical consciousness that our index measures. Following standard practices~\cite{boateng2018best},
\color{black}
we retained factors with eigenvalues greater than $1.0$ 
(indicating factors that explain more variance than individual items)
\color{black}
and items with factor loadings of at least $0.4$
(representing strong correlations between items and their factors).
\color{black}
We also applied a promax rotation to allow for correlations between factors \cite{corner2009choosing}. We performed EFA separately for each hypothesized construct as recommended by psychometric scale development practices \cite{comrey2013first}.

\subsection{Scale Evaluation}

\subsubsection{Tests of Dimensionality}

To validate the factor structure found from EFA, we conducted confirmatory factor analysis (CFA) with the second sample. We assessed model fit using established metrics: These included the Root Mean Square Error of Approximation (RMSEA < .06), Tucker Lewis Index (TLI > .90), Comparative Fit Index (CFI > .90), and Standardized Root Mean Square Residual (SRMR < .08) \cite{boateng2018best}.

\subsubsection{Tests of Reliability and Validity}

We evaluated reliability through Cronbach's alpha to assess internal consistency across factors. 
For validity, \citeauthor{padiyath2025development} gathered evidence on content validity through expert review and theoretical grounding \cite{padiyath2025development}, and in this study, we gathered evidence for construct validity through CFA testing of factor structures hypothesized during EFA. Following \citet{raykov2011introduction}, we assessed construct validity through a two-phase process: First, we examined correlations between constructs expected to be related, looking for correlation coefficients greater than +.5 or less than -.5. Second, we examined correlations between theoretically unrelated constructs, expecting correlation coefficients closer to 0. 

To assess predictive criterion validity (the ability of the survey to predict ethical behavior \cite{csedresearchMeasuringValidity}), we asked additional questions about respondents' workplace experience. We first asked if the respondent had industry experience as a computer scientist or software developer, including co-ops and internships (Yes/No). For those with industry experience, we asked whether they had encountered ethical concerns about products or practices in their workplace (Yes/No). Participants who said they had encountered ethical concerns were then asked if they took any action to respond to those issues (Yes/No). Based on how developers typically handle ethical issues as found by \citet{widder2023s}, we provided examples of what we meant by ``taking action'' such as discussing concerns with colleagues, human resources, or management; refusing to work on problematic projects; and/or notifying people and peers outside of their organization \cite{widder2023s}. We then conducted linear regressions to understand the relationship between our measured constructs and noticing/acting on ethical issues, controlling for student status, ethnicity, gender, and age. We excluded respondents with missing demographic information.
Given no established benchmark measure, we did not assess concurrent criterion validity, aligning with common practice in similar contexts \cite{boateng2018best}.

\subsection{Computing Ethics Education Effects on Critical Reflection and Agency}

During the CFA study administration, participants also were asked whether they had completed a computing ethics or professionalism course (Yes/No). We conducted a linear regression analysis \cite{draper1998applied} to determine whether these ethics training experiences related to developers' sociotechnical awareness and agency, as measured by our instrument. The regressions predicted each scale factor using ethics course completion as the predictor, while controlling for demographic variables: student status, ethnicity, gender, and age. We excluded respondents with missing demographic information. All regressions were conducted using StataSE 18. Each regression and its assumptions were then validated with professional statistics consultants who are experts in regression analysis.

\subsection{Limitations}

We recognize the inherent challenges in using quantitative measures to capture complex social phenomena, a perspective known as CritQuant \cite{diemer2023illustrating, frisby2024critical}. Quantitative scales, by their nature, reduce multifaceted concepts like critical consciousness and ethical awareness into simplified numeric representations. This reduction can potentially reinforce existing power structures by determining what gets measured and how. While we believe our index offers valuable insights, we must acknowledge that the act of measurement itself shapes how we understand and discuss critical consciousness in computing. The rapidly evolving nature of computing ethics also raises questions about the temporal stability of the scale's constructs as discussed in \citet{padiyath2025development}: what we measure as critical consciousness today may shift as new ethical challenges emerge in computing.

Our validation study also faced several methodological limitations. The sample demographics, while diverse in some aspects, came primarily from United States-based computing programs and may not capture perspectives from other cultural and educational contexts. While this U.S. focus allowed us to control for variations in educational systems and professional contexts that might affect our validation process, it limits the generalizability of our findings to other cultural contexts. Additionally, our validation relied heavily on self-reported measures, which can be subject to social desirability bias -- the tendency of survey respondents to answer in a manner viewed favorably by others \cite{chung2003exploring} -- especially when dealing with issues of ethics \cite{tan2021social}.

While our binary outcome variables (noticing and acting on ethical issues) enabled clearer initial validation of the relationship between critical consciousness and ethical behavior, we oversimplified what we know is a more nuanced spectrum of ethical awareness and action \cite{widder2023s}. Future work could explore the relationship between critical consciousness and more sophisticated outcome measures. Additionally, while we controlled for common demographic variables, other factors could influence both our predictor variables and outcomes, potentially causing omitted variable bias.

Our measurement of computing ethics education also relied on a binary question about completing a dedicated computing ethics or professionalism course. This approach may not capture the effect of newer curricula that embed ethics throughout their curriculum in addition to or instead of delivering it in a standalone format~\cite{grosz2019embedded}. Programs with embedded ethics might produce different patterns in critical reflection and agency compared to a standalone ethics course-based education.
\color{black}

While we found associations between critical consciousness measures and ethical behavior, we cannot definitively establish whether higher critical consciousness leads to increased ethical action or if those who take ethical action develop stronger critical consciousness. Additionally, our linear models assumed linear relationships between predictors and outcomes. While this is standard practice for exploratory research, it may not fully capture more complex interaction effects between different dimensions of critical consciousness.

\section{Results}

\subsection{RQ1: Scale Development}

\subsubsection{Sampling and Survey Administration}

Our first survey administration yielded 500 responses through Prolific, with 474 participants successfully completing all three attention check questions. The sample contained ranges across multiple demographics: participants averaged 29.06 years of age (SD = 9.97), with gender representation including men (68\%), women (29\%), non-binary individuals (3\%), and those identifying as both non-binary and a man (.2\%). Racial diversity was also evident, with participants identifying as White (49\%), Asian (18\%), Mixed race (15\%), Black (14\%), and Other (3.59\%). A majority (57\%) were current undergraduate students.

Our second sample also yielded 500 responses, with 464 participants successfully completing all three attention check questions. The sample's demographics were similar to the first sample: participants averaged 29.08 years of age (SD = 9.91); gender representation included men (64\%), women (31\%), non-binary individuals (4\%), genderqueer (.2\%), those identifying as both a man and a woman (.6\%), and those identifying as both non-binary and a woman (.4\%); Participants identified as White (48\%), Asian (15\%), Mixed race (13\%), Black (19\%), and Other (5\%); and 58\% were current undergraduate students.

\subsubsection{Extraction of Factors}

Prior to conducting factor analysis, we verified the appropriateness of our data for factor analysis. The Kaiser-Meyer-Olkin measure of sampling adequacy exceeded the threshold for ``meritorious'' (KMO~>~.8) \cite{kaiser1974little}, while the Bartlett test of sphericity showed significant results (p < .0001) \cite{boateng2018best}, confirming sufficient inter-item correlations for factor analysis. The first construct, ``Recognizing computing/data embeds values/power,'' revealed a factor structure that differed from our hypothesized structure: Several reverse-coded items from the hypothesized subscales merged into a single factor. We labeled this factor ``Techno-solutionism,'' reflecting the belief that technological solutions alone can solve all problems \cite{morozov2013save, toyama2015geek}. The remaining items from the ``Data has limits'' subscale did not achieve significant factor loadings and were thus removed from the validation study. The remaining items from ``Centering power in ethics'' were related to valuing the perspectives of marginalized communities, and were thus renamed as such. The complete factor structure from EFA is shown in Appendix \ref{appendix:efa_loadings}.

\subsection{RQ2: Scale Evaluation}

\subsubsection{Tests of Dimensionality}

To verify the factor structure identified through EFA, we conducted two separate confirmatory factor analyses (CFA) using the second administered sample. Our initial CFA tested a model with three factors for the first construct, one factor for the second construct, and two factors for the third construct. However, after observing an extremely low Cronbach's alpha ($\alpha$ = .38) for the ``computing has limits'' factor, we conducted a second CFA excluding this factor. Results support the latter model, as it scored adequately on the following: Root Mean Square Error of Approximation (RMSEA < .06) \cite{hu1999cutoff}, Tucker Lewis Index (TLI > .90) \cite{byrne1994structural}, Comparative Fit Index (CFI > .90) \cite{byrne1994structural}, and Standardized Root Mean Square Residual (SRMR < .08) \cite{hu1999cutoff}. The factor loadings of all items are above the threshold of .55 defined as ``good'' by Comrey and Lee \cite{comrey2013first}. All factor loadings can be found in Appendix \ref{appendix:cfa_loadings}.

While we excluded the ``computing has limits'' factor from our final CFA model, we decided to keep the items in the final index. Although they cannot function as a reliable subscale with psychometric properties, since they demonstrated face validity and content validity, we believe they still provide valid reference points for assessing students' views as standalone items to be compared to each other, rather than as a unified construct.

\subsubsection{Tests of Reliability}

Internal consistency demonstrated good reliability across most subscales, with Cronbach's alpha coefficients ranging from .63 to .89. The techno-solutionism ($\alpha$ = .79), valuing ethics training ($\alpha$ = .89), personal effectiveness ($\alpha$ = .89), and system responsiveness ($\alpha$ = .84) subscales all exceeded the recommended threshold of .7 \cite{nunnally1967assessment}. While the valuing marginalized perspectives subscale ($\alpha$ = .63) fell below this threshold, it meets the .6 criterion for exploratory research \cite{nunnally1967assessment}, suggesting the subscale provides useful preliminary insights while indicating room for future improvement.

\subsubsection{Tests of Validity}

Our construct validity assessment examined theoretically predicted relationships between factors (see Table \ref{tab:corrs}). Personal Effectiveness and System Responsiveness showed a strong positive correlation (r = .62, p < .0001), supporting our framework's operationalization of these as different facets of agency in computing ethics. This pattern aligns with findings in critical consciousness and political efficacy research \cite{diemer2017development, diemer2022development, watts2011critical}. 

The significant positive correlation between valuing ethics training and valuing marginalized perspectives (r = .53, p < .0001) aligns with critical consciousness theory, which posits that critical reflection encompasses both recognition of systemic inequities and a commitment to addressing them. Valuing ethics training showed moderate positive correlations with both Personal Effectiveness (r = .37, p < .0001) and System Responsiveness (r = .28, p < .0001), suggesting these constructs capture different aspects of critically conscious computing.

\begin{table*}[]
    \centering
    \begin{tabular}{ccccccc}
    \toprule
        & \makecell{Valuing \\ Marginalized \\ Perspectives} & \makecell{Techno-\\Solutionism} & \makecell{Valuing \\ Ethics Ed} & \makecell{Valuing \\ Technical Ed} & \makecell{Personal\\ Effectiveness} & \makecell{System \\Responsiveness}\\
        \midrule
        \makecell{Valuing \\ Marginalized \\ Perspectives} & 1.0 \smallskip \\
        \makecell{Techno-\\Solutionism} & .00 & 1.0 \smallskip \\
        \makecell{Valuing \\ Ethics Training} & \textbf{.53} & .03 & 1.0 \smallskip \\
        \makecell{Valuing \\ Technical Training} & \textbf{.21} & .04 & \textbf{.44} & 1.0 \smallskip \\
        \makecell{Personal\\ Effectiveness} &\textbf{ .24 }& \textbf{.20 }& \textbf{.37 }& \textbf{.26} & 1.0\smallskip \\
        \makecell{System \\Responsiveness} & \textbf{.13} & \textbf{.44 }&\textbf{ .28} & \textbf{.16} & \textbf{.62} & 1.0\smallskip \\
        \bottomrule
    \end{tabular}
    \caption{Factor Correlations in CFA Study. Bolded correlations indicate significance ($p$ < .05).}
    \label{tab:corrs}
\end{table*}

Interestingly, the techno-solutionism factor showed moderate positive correlations with both agency constructs (Personal Effectiveness: r = .20, p < .0001; System Responsiveness: r = .44, p < .0001). Without prior theory about these relationships, we cannot interpret these correlations as evidence for or against construct validity. However, they suggest an interesting pattern: individuals holding stronger techno-solutionist beliefs report higher levels of agency. This could reflect either a greater confidence in technological solutions to systemic ethical problems or perhaps simply a generally optimistic outlook.

To assess predictive criterion validity, we examined whether our instrument's measures were associated with participants' real-world ethical behavior -- specifically, identifying and responding to ethical issues. We conducted two sets of linear regressions: one set exploring relationships with noticing ethical issues and another with taking action. Given the multiple comparisons problem, we applied the Bonferroni correction~\cite{dunn1961multiple}, adjusting our significance level for the first set of linear regressions to $\alpha$ = 0.05/3 = 0.017 and for our second set of linear regressions to $\alpha$ = 0.05/2 = 0.025.

\begin{table*}[]
    \centering
    \begin{tabular}{lccccc}
    \toprule
         & \makecell{Valuing \\ Ethics \\ Training} & \makecell{Techno-\\solutionism} & \makecell{Valuing \\ Marginalized \\ Perspectives} & \makecell{Personal\\Effectiveness} & \makecell{System\\Responsiveness} \\
    \midrule
       \makecell[l]{Noticed an Ethical Issue}  & \makecell{.13*\\(.12)}& \makecell{.04\\(.14)} & \makecell{.12*\\(.14)} & \multicolumn{2}{c}{---} \\
       \makecell[l]{Acting on an Ethical Issue}  & \multicolumn{3}{c}{---} & \makecell{.16*\\(.20)}& \makecell{.21*\\(.24)} \\
       Student & \makecell{-.01\\(.12)}& \makecell{-.05\\(.14)} & \makecell{.06\\(.14)} & \makecell{-.14\\(.16)}& \makecell{-.05\\(.19)} \\
       Ethnicity (White)  & &  &  & & \\
       \hspace{2mm} Asian   & \makecell{.03\\(.19)} & \makecell{-.03\\(.22)}& \makecell{.12\\(.22)} & \makecell{-.01\\(.26)} & \makecell{.14\\(.32)}\\
       \hspace{2mm} Black   &\makecell{.19**\\(.16)} & \makecell{.19**\\(.19)}& \makecell{.21**\\(.19)} &\makecell{.12\\(.20)} & \makecell{.28**\\(.24)}\\
       \hspace{2mm} Mixed   & \makecell{.05\\(.18)}& \makecell{.10\\(.21)} & \makecell{.10\\(.22)} & \makecell{.01\\(.23)}& \makecell{.13\\(.28)} \\
       \hspace{2mm} Other   & \makecell{.02\\(.31)}& \makecell{.09\\(.36)}& \makecell{.03\\(.36)} & \makecell{-.18\\(.53)}& \makecell{-.07\\(.64)}\\
       Age  & \makecell{.11\\(.01)} & \makecell{-.03\\(.01)}& \makecell{.09\\(.01)} & \makecell{.19*\\(.01)} & \makecell{-.08\\(.01)}\\
       Gender (Man) &  &  &  & & \\
       \hspace{2mm} Man, Woman & \makecell{.00\\(.58)} & \makecell{.02\\(.67)}& \makecell{.06\\(.68)} & \makecell{.08\\(.53)} & \makecell{.04\\(.64)}\\
       \hspace{2mm} Non-binary & \makecell{.14*\\(.31)} & \makecell{-.12*\\(.36)}& \makecell{.16**\\(.36)} & \makecell{-.02\\(.46)} & \makecell{-.14\\(.56)}\\
       \hspace{2mm} Woman & \makecell{.10\\(.14)} & \makecell{-.08\\(.16)} & \makecell{.16**\\(.16)} & \makecell{-.01\\(.17)} & \makecell{-.02\\(.21)} \\
       \hspace{2mm} \makecell[l]{Woman,\\Non-binary} & \makecell{.04\\(.99)}& \makecell{-.05\\(1.14)}& \makecell{.11\\(1.16)} & \multicolumn{2}{c}{---} \\     
    \midrule
    N & 278 & 278 & 278 & 132 & 132 \\
    $F$ & 2.33** & 2.24** & 3.18*** & 2.15* & 2.65** \\
    $R^2$ & .09 & .08 & .12 & .15 & .18 \\
    \bottomrule
    \end{tabular}
    \caption{Regression Analysis Results: Factors Associated with Noticing and Acting on Ethical Issues in the Workplace. Standardized coefficients ($\beta$) with standard errors in parentheses. *$p$ < .05, **$p$ < .01, ***$p$ < .001. The first three columns represent factors associated with noticing ethical issues, while the last two columns represent factors associated with taking action on ethical issues.}
    \label{tab:LR_criterion_combined}
\end{table*}

For noticing ethical issues, we conducted linear regressions with valuing ethics training, valuing marginalized perspectives, or techno-solutionism, since our theory predicts these as related to noticing ethical issues (see Table \ref{tab:LR_criterion_combined}). Valuing ethical training and valuing marginalized perspectives showed a significant association after controlling for student status, ethnicity, age, and gender. Although we expected a negative relationship between noticing ethical issues and techno-solutionism, there was a non-significant relationship with techno-solutionism, suggesting either weak associations with ethical awareness, or a more complex relationship that our model could not capture.

For acting on ethical issues, we tested if specifically personal effectiveness or system responsiveness showed a relationship with ethical action (see Table \ref{tab:LR_criterion_combined}). Both personal effectiveness and system responsiveness showed a strong relationship with acting on ethical issues, aligning with our theory.

These significant relationships between our instruments' measures and ethical behavior in the workplace provide evidence for the scale's predictive criterion validity.

\subsection{Final Index}

After conducting exploratory and confirmatory factor analysis, the final Critical Reflection and Agency in Computing Index consists of 29 likert-scale items distributed across two main constructs: Critical Reflection (22 items) in Tables \ref{tab:critical_reflection1} and \ref{tab:critical_reflection2} and Critical Agency (9 items) in Table \ref{tab:critical_agency}. Each construct is further divided into sub-constructs to capture different aspects of critical consciousness in computing. We recommend administering the index with a 4- or 6-item Likert-scale format, with respondents indicating their level of agreement or disagreement with each statement.

While we excluded the ``computing has limits'' factor from our final CFA model, we made the decision to retain two of its items: ``Computing should inform, not replace, human decision-making'' and ``We should prioritize computational solutions over human judgment'' (R). Though these items cannot form a reliable subscale, they demonstrated face and content validity. Thus we believe they are still useful as reference points for understanding students' views on computing's role in decision-making.

\begin{table*}
    \centering
    \begin{tabular}{p{2.5cm}p{2.75cm}p{8.25cm}}
    \toprule
    \multicolumn{3}{p{13.5cm}}{\textbf{Question Wording}: Computing technologies have wide-ranging impacts on society. Please indicate the extent to which you agree with the following statements:} \\
    \midrule
        \textbf{Construct} & \textbf{Operationalization} & \textbf{Item} \\
    \midrule
        \multirow{8}{8em}{Recognizing computing/data embeds values/power:} & & Computing should inform, not replace, human decision-making. \\
        & & We should prioritize computational solutions over human judgment. (R) \\
        & \textit{Techno-solutionism} & With enough resources, computing technologies can solve any problem. (R) \\
        & \textit{Techno-solutionism} & Datasets that are large enough can overcome any bias in collection. (R) \\
        & \textit{Techno-solutionism} & Biases in datasets can always be corrected with the right techniques. (R) \\
        & \textit{Techno-solutionism} & Computing technologies benefit everyone equally. (R) \\
        & \makecell[l]{\textit{Valuing Marginalized}\\\textit{Perspectives}} & Considering issues of social justice should be a fundamental consideration in the design and development of any computing system. \\
        & \makecell[l]{\textit{Valuing Marginalized}\\\textit{Perspectives}} & Developing computer software for public use requires input from marginalized groups. \\
    \bottomrule
    \end{tabular}
    \caption{Critical Reflection Subscale: Recognizing computing/data embeds values/power. (R) denotes a reverse-coded item.}
    \label{tab:critical_reflection1}
\end{table*}

\begin{table*}
    \centering
    \begin{tabular}{p{2.5cm}p{3.5cm}p{7.5cm}}
    \toprule
    \multicolumn{3}{p{13.5cm}}{\textbf{Question Wording}: Different professions require different training. Please indicate the extent to which you agree that the following should be part of training for every software engineer:} \\
    \midrule
        \textbf{Construct} & \textbf{Operationalization}  & \textbf{Item} \\
    \midrule
    \multirow{14}{8em}{Recognizing computing training should include more explicit ethics and social impact discussions:} & \textit{Valuing Ethics Training} & The social impacts of software. \\
     & \textit{Valuing Ethics Training} & The environmental impacts of software. \\
     & \textit{Valuing Ethics Training} & Legal considerations in software development. \\
     & \textit{Valuing Ethics Training} & Ethical implications of topics being studied. \\
     & \textit{Valuing Ethics Training} & Collaborating on software development projects with local community groups. \\
     & \textit{Valuing Ethics Training} & Guidelines for discussing ethical issues with others. \\
     & \textit{Valuing Ethics Training} & A software development code of ethics. \\
     & \textit{Valuing Technical Training} & Computer architectures. (0) \\
     & \textit{Valuing Technical Training} & Databases. (0) \\
     & \textit{Valuing Technical Training} & Technical programming skills. (0) \\
     & \textit{Valuing Technical Training} & Software quality assurance and testing. (0) \\
     & \textit{Valuing Technical Training} & Computer science theory and algorithms. (0) \\
     & \textit{Valuing Technical Training} & Identifying requirements to build software. (0) \\
     & \textit{Valuing Technical Training} & Data structures. (0) \\
    \bottomrule
    \end{tabular}
    \caption{Critical Reflection Subscale: Recognizing computing training should include more explicit ethics and social impact discussions.}
    \label{tab:critical_reflection2}
\end{table*}

\begin{table*}
    \centering
    \begin{tabular}{p{3cm}p{3cm}p{7.5cm}}
    \toprule
    \multicolumn{3}{p{13.5cm}}{\textbf{Question Wording}: Please indicate the extent to which you agree with the following:		} \\
    \midrule
        \textbf{Construct} & \textbf{Operationalization} & \textbf{Item} \\
    \midrule
    \multirow{10}{10em}{Belief that computing professionals have agency:} & \textit{Personal effectiveness} & I have a good understanding of the important ethical and social impacts to consider when developing software. \\
    & \textit{Personal effectiveness} & I am able to participate in discussions about ethics and social impacts of computing. \\
    & \textit{Personal effectiveness} & I am confident in my own ability to uphold ethical conduct in software development. \\
    & \textit{Personal effectiveness} & I am better informed about the ethics and societal impacts of technology than most of my software developer peers. \\
    & \textit{Personal effectiveness} & When working on computing projects with others, I could effectively voice my perspectives on ethical issues. \\
    & \textit{System responsiveness} & There are processes within workplaces to handle reported ethical computing violations or concerns. \\
    & \textit{System responsiveness} & When I talk about ethical computing issues, my peers usually pay attention. \\
    & \textit{System responsiveness} & Software development professionals are allowed to have a say about ethical computing concerns at their workplaces. \\
    & \textit{System responsiveness} & When ethical computing concerns are raised by employees, workplaces are responsive to addressing these concerns. \\
    \bottomrule
    \end{tabular}
    \caption{Critical Agency Subscales: Personal Effectiveness and System Responsiveness}
    \label{tab:critical_agency}
\end{table*}

\subsection{RQ3: Computing Ethics Education Effects on Critical Reflection and Agency}

\subsubsection{Respondents Attitudes towards Critical Reflection and Agency in Computing.}

Appendix \ref{appendix:sample_stats} provides the means and standard deviations of the measures used for the respondents in the CFA sample. More than 75\% of respondents reported at least somewhat valuing computing ethics in their training ($\mu$ = 4.63, $\sigma$ = .98 on 1 = strongly disagree to 6 = strongly agree scales). However, when compared to valuing a technical computing education, more than 90\% of respondents reported at least somewhat valuing technical training ($\mu$ = 5.20, $\sigma$ = .74). A paired t-test shows respondents, on average, significantly valuing technical training more than computing ethics training ($t$(473) = -13.25, $p$ < .0001).

\subsubsection{Computing Ethics Training Effects on Attitudes towards Critical Reflection and Agency}

To understand if computing ethics courses and training have a relationship with the constructs in our study, we use it as a predictor for our regressions predicting each of our constructs. Based on our regressions (see Table \ref{tab:cfa_regressions_ethics_class}), we find that computing ethics training has a statistically significant positive relationship with valuing ethics training, personal effectiveness and system-responsiveness. Interestingly, it also has a significantly slightly positive relationship with techno-solutionism.

\begin{table*}[]
    \centering
    \begin{tabular}{lcccccc}
    \toprule
         & \makecell{Valuing \\ Ethics \\ Training} & \makecell{Valuing \\ Technical \\ Training} & \makecell{Techno-\\solutionism} & \makecell{Valuing \\ Marginalized \\ Perspectives} & \makecell{Personal\\Effectiveness} & \makecell{System\\Responsiveness} \\
    \midrule
       \makecell[l]{Computing\\Ethics Training}  & \makecell{.11*\\(.09)} & \makecell{-.01\\(.07)} & \makecell{.17***\\(.10)} & \makecell{.08\\(.11)} & \makecell{.30***\\(.10)} & \makecell{.27***\\(.09)}\\
       Student & \makecell{-.01\\(.09)} & \makecell{-.09\\(.07)} & \makecell{-.06\\(.10)} & \makecell{.04\\(.11)} & \makecell{-.07\\(.10)} & \makecell{-.04\\(.10)}\\
       Ethnicity (White)  & &  &  &  &  & \\
       \hspace{2mm} Asian   & \makecell{.03\\(.14)}& \makecell{-.01\\(.10)} & \makecell{.03\\(.15)} & \makecell{.13**\\(.16)} & \makecell{-.02\\(.14)} & \makecell{.04\\(.14)}\\
       \hspace{2mm} Black   &\makecell{.15**\\(.12)} & \makecell{-.07\\(.09)} & \makecell{.15**\\(.14)} & \makecell{.18***\\(.15)} & \makecell{.04\\(.13)} & \makecell{.09\\.13}\\
       \hspace{2mm} Mixed   & \makecell{.07\\(.14)}& \makecell{-.09\\(.11)} & \makecell{.10*\\(.16)} & \makecell{.11*\\(.17)} & \makecell{.10*\\(.15)} & \makecell{.05\\(.15)}\\
       \hspace{2mm} Other   & \makecell{.01\\(.22)}& \makecell{-.05\\(.16)} & \makecell{.05\\(.24)} & \makecell{.02\\(.26)} & \makecell{-.06\\(.23)} & \makecell{-.07\\(.22)}\\
       Age  & \makecell{.07\\(.00)} & \makecell{.03\\(.00)} & \makecell{-.00\\(.01)} & \makecell{.08\\(.01)} & \makecell{.15***\\(.01)} & \makecell{.03\\(.00)}\\
       Gender (Man) &  &  &  &  &  & \\
       \hspace{2mm} Genderqueer & \makecell{.06\\(.97)} & \makecell{.01\\(.74)} & \makecell{-.06\\(1.10)} & \makecell{.09\\(1.16)} & \makecell{-.01\\(1.02)} & \makecell{-.09*\\(.99)}\\
       \hspace{2mm} Man, Woman & \makecell{.00\\(.57)} & \makecell{.03\\(.43)} & \makecell{.03\\(.63)} & \makecell{.06\\(.68)} & \makecell{.06\\(.60)} & \makecell{.06\\(.58)}\\
       \hspace{2mm} Non-binary & \makecell{.15***\\(.24)} & \makecell{.06\\(.18)} & \makecell{-.12**\\(.26)} & \makecell{.18***\\(.28)} & \makecell{-.07\\(.25)} & \makecell{-.12**\\(.24)}\\
       \hspace{2mm} Woman & \makecell{.12**\\(.10)} & \makecell{-.00\\(.07)} & \makecell{-.11*\\(.11)} & \makecell{.20***\\(.12)} & \makecell{-.02\\(.10)} & \makecell{-.03\\(.10)}\\
       \hspace{2mm} \makecell[l]{Woman,\\Non-binary} & \makecell{.05\\(.69)} & \makecell{-.01\\(.52)} & \makecell{-.05\\(.77)} & \makecell{.07\\(.82)} & \makecell{.03\\(.73)} & \makecell{-.01\\(.70)}\\     
    \midrule
    N & 472 & 472 & 472 & 472 & 472 & 472\\
    $F$ & 2.90*** & .38 & 4.40*** & 4.52*** & 6.46*** & 5.56***\\
    $R^2$ & .07 & .03 & .10 & .11 & .14 & .13\\
    \bottomrule
    \end{tabular}
    \caption{Regression Analysis Results: Impact of Computing Ethics Training on Critical Reflection and Agency Factors. Standardized coefficients ($\beta$) with standard errors in parentheses. *$p$ < .05, **$p$ < .01, ***$p$ < .001}
    \label{tab:cfa_regressions_ethics_class}
\end{table*}

\section{Discussion}

\subsection{Representing Critical Consciousness in Computing}

While our data validates the theoretical framework of \textit{Critically Conscious Computing}, it also exposes a tension that \citet{ko2020time} described in their call for a more critical CS education -- the challenge of helping students recognize both the power \textit{and} limitations of technology. The emergence of techno-solutionism as a distinct construct, with its positive correlations with agency measures (r = .20 for personal effectiveness; r = .44 for system responsiveness) suggests that students' confidence in addressing ethical issues may paradoxically stem from their faith in technical solutions. This is precisely the mindset that scholars like \citet{ko2020time} and \citet{morozov2013save} warn against when advocating for criticality in computing.
However, this is one interpretation of the data, future work could explore whether this correlation reflects a causal relationship or a more nuanced attitude toward technology that our measure couldn't capture.
\color{black}

Our data also presents evidence for the validity of our framework for critical agency, as personal effectiveness and system responsiveness emerged as distinct but related dimensions. Unlike general political efficacy measures focused on governmental change, this framework captures the specific tensions computing professionals navigate between individual capacity and organizational constraints that \citeauthor{widder2023s} identified in their study of industry practices \cite{widder2023s}. The strong reliability of these measures provides quantitative evidence supporting theoretical models of how students develop confidence in addressing ethical challenges within computing contexts.

Another interesting insight came from what couldn't be measured reliably. Despite \citeauthor{padiyath2025development}'s careful item development, questions about computing and data limitations did not form a reliable subscale,
meaning these items didn't consistently measure the same underlying concept across participants' responses.
\color{black}
This leads to a open question in computing ethics education: How do we best assess students' understanding of the limits of technology? Our findings show that recognizing computing's limits isn't simply the opposite of techno-solutionism - otherwise these items would have loaded negatively onto that factor. This might suggest that students' conceptualization of the limits of computing may be more complex than a straightforward rejection of techno-solutionism. However, \citeauthor{padiyath2025development}'s scale development process didn't succeed in reliably measuring this dimension of critical reflection.

\subsection{The Impact of Computing Ethics Education}

One of our most exciting findings addresses a fundamental question in computing ethics education: Does ethics training actually shift students' ethical orientations? Our results provide empirical evidence that computing ethics education is associated with measurable differences in critical consciousness. Students who completed ethics courses scored significantly higher on key dimensions of our index, particularly in personal effectiveness ($\beta$ = .30, p < .001) and system responsiveness ($\beta$ = .27, p < .001), even when controlling for demographics and student status. This suggests that ethics education, 
even without knowing the specific pedagogical approaches or formats used (ranging from dedicated ethics courses to brief modules on professionalism),
\color{black}
is developing the very traits our profession has theorized as necessary for ethical behavior in practice.

While ethics training showed a modest positive correlation with valuing ethics training ($\beta$ = .11, p < .05), its relationship with valuing marginalized perspectives was not quite significant ($\beta$ = .08, p = 0.07). This suggests that our ethics education approaches may be more successful at conveying the importance of ethical considerations broadly than at fostering specific appreciation for diverse perspectives and social justice concerns -- which \citet{washington2020twice} and \citet{vakil2018ethics} identify as essential for critical computing education. Additionally, ethics training showed positive correlations not only with critical agency but also with techno-solutionism ($\beta$ = .17, p < .001). We may be inadvertently reinforcing the idea that ethical challenges are primarily technical puzzles awaiting technical solutions.

Our results also quantitatively confirm what we qualitatively know as a disparity in how students value technical versus ethical training, as technical training was consistently rated as more important (t(473) = -13.25, p < .0001). This echoes \citeauthor{cech2014culture} findings on a ``culture of disengagement'' in engineering education and quantitatively confirms what \citeauthor{ko2020time} and others \cite{brown2023designing, fiesler2021integrating, vakil2018ethics, padiyath2024undergraduate, garrett2020more} have identified as computing education's core challenge: integrating ethical considerations as fundamental rather than an add-on to computing education. Our index now provides educators and researchers with a concrete way to systematically measure progress toward closing this gap.

\subsection{Implications for Computing Ethics Education and Assessment}

In validating the Critical Reflection and Agency in Computing Index, our field now has a measurement tool that addresses a fundamental gap in computing ethics education: the ability to systematically assess ethical development. This index provides a theoretically-grounded framework with strong psychometric properties, addressing the limitations identified by \citet{brown2024teaching} and \citet{padiyath2024realist} of previous works' use of intervention-specific surveys without evidence for validity.

By using the index prior to ethics courses and modules, educators can design targeted interventions based on specific class profiles rather than assume one-size-fits-all approaches. As \citeauthor{padiyath2024realist} found, this tailoring might be critical for an intervention's success \cite{padiyath2024realist}. Use of the index can reveal whether students struggle with valuing marginalized perspectives, believing in their capacity to enact change, or maintaining critical perspectives on technology's capabilities, each requiring different pedagogical responses. With a paired post-survey and qualitative approaches, educators can measure specific impacts of their interventions on students' critical consciousness. 
While we recommend using the complete validated instrument when possible, researchers may use specific subscales when only certain factors are relevant to their research questions. However, we warn against modifying item wording or structure without conducting additional validation.
\color{black}

Looking forward, this index enables direct comparison across different interventions, institutions, and student populations. This standardization can help establish evidence-based best practices in our field. The index highlights key areas where computing education needs to evolve: balancing technical skills with critical consciousness, understanding both the power and limitations of technical solutions, and preparing students to navigate ethical challenges within workplaces. By integrating these elements into computing curricula, we can prepare computing professionals to better address the ethical demands of our complex sociotechnical environments.

\section{Conclusion}

Computing's influence across society demands that we understand how students' develop ethical agency to ensure they are responsible. By validating the Critical Reflection and Agency in Computing Index, our field now has a theoretically-grounded instrument that reveals both the impacts and gaps in current educational approaches to computing ethics. Our findings indicate that while ethics courses foster some dimensions of critical reflection and agency, they may simultaneously reinforce techno-solutionist mindsets. This validated index enables the computing education community to monitor students' ethical development across institutions and pedagogical approaches, identify class-specific intervention needs, and build evidence-based best practices. As computing continues to shape our world, our research provides a foundational tool for preparing future professionals with the critical consciousness needed to responsibly shape computing technologies.

\bibliographystyle{ACM-Reference-Format}
\bibliography{sample-base}

\clearpage

\appendix

\section{Exploratory Factor Analysis Loadings}
\label{appendix:efa_loadings}

\begin{table*}[h!]
    \centering
    \begin{tabular}{p{10cm}ccc}
    \toprule
       \textbf{Item} & \textbf{Factor 1} & \textbf{Factor 2} & \textbf{Factor 3} \\
    \midrule
        Computing should inform, not replace, human decision-making.& -.0437  &  \textbf{.4038}  &  .0805  \\
        It’s important to explore if a non-computing solution would be the best approach for a given problem. & -.0977  &  .1243  &  .0425  \\
        With enough resources, computing technologies can solve any problem. (R) &  \textbf{.5983}  &  .1469  &  .0018  \\
        We should prioritize computational solutions over human judgment. (R) &  .2756  &  \textbf{.4668}  & -.0256  \\
        Data reflects the past and does not fully capture current reality. & .0783  & -.1565  &  .0173  \\
        Using data requires considering the data's limitations. &  .0848  & -.0410  &  .0108  \\
        Datasets that are large enough can overcome any bias in collection. (R)  &  \textbf{.7527}  & -.0103  &  .0457  \\   
        Biases in datasets can always be corrected with the right techniques. (R) &  \textbf{.6880}  & -.0923  &  .0040  \\
        Considering issues of social justice should be a fundamental consideration in the design and development of any computing system. & -.0094  &  .0116  &  \textbf{.6574} \\
        Developing computer software for public use requires input from marginalized groups.  & .0402  & -.0059  &  \textbf{.6602}  \\
        Computing technologies benefit everyone equally. (R) &  \textbf{.5884}  &  .0557  & -.0436  \\
        Ethics discussions in computing should only involve computer scientists. (R)  & \textbf{.4095}  &  .1404  & -.0167  \\
    \bottomrule
    \end{tabular}
    \caption{Exploratory Factor Analysis for the Critical Reflection Subscale: Recognizing computing/data embeds values/power. (R) denotes a reverse-coded item. Bolded factor loadings indicate loadings above the recommended minimum of .4.}
    \label{tab:EFA_computing_power}
\end{table*}

\begin{table*}[h!]
    \centering
    \begin{tabular}{p{10.5cm}cc}
    \toprule
        \textbf{Item} & \textbf{Factor 1} & \textbf{Factor 2} \\
    \midrule
     The social impacts of software. &  .1112 & \textbf{ .8025}\\
     The environmental impacts of software. & -.1501 & \textbf{ .6726}\\
     Legal considerations in software development. &  .1215 & \textbf{ .4211}\\
     Engaging with stakeholders affected by software projects. & -.0766 &  .1688\\
     Ethical implications of topics being studied. &  .0524 & \textbf{ .8231}\\
     Collaborating on software development projects with local community groups. & -.0986 & \textbf{ .4502}\\
     Software development professional responsibilities. &  .3859 &  .1862\\
     Guidelines for discussing ethical issues with others. & -.1041 & \textbf{ .7877}\\
     A software development code of ethics. &  .2095 & \textbf{ .5887}\\
     Discrete mathematics. &  .2718  & -.0006\\
     Computer architectures. & \textbf{ .5230}  &  .0455\\
     Databases. & \textbf{ .7285}  & -.0642\\
     Technical programming skills. &  \textbf{ .7346} &  .0217\\
     Delivering software projects on-time and within budget. &  .1766 & -.0924\\
     Software quality assurance and testing. &  \textbf{ .5257} & -.0011\\
     Computer science theory and algorithms. &  \textbf{ .5187} &  .0094\\
     Identifying requirements to build software. &  \textbf{ .5236} &  .0068\\
     Data structures. &\textbf{ .8303}  & -.0305 \\
        \bottomrule
    \end{tabular}
    \caption{Exploratory Factor Analysis for the Critical Reflection Subscale: Recognizing computing training should include more explicit ethics and social impact discussions. Bolded factor loadings indicate loadings above the recommended minimum of .4.}
    \label{tab:EFA_critical_reflection2}
\end{table*}

\begin{table*}[h!]
    \centering
    \begin{tabular}{p{10.5cm}cc}
    \toprule
    \textbf{Item} & \textbf{Factor 1} & \textbf{Factor 2} \\
    \midrule
    I have a good understanding of the important ethical and social impacts to consider when developing software.&  \textbf{.7542}  &  .0290 \\ 
    I am able to participate in discussions about ethics and social impacts of computing.&  \textbf{.7976}  & -.0697 \\ 
    I am confident in my own ability to uphold ethical conduct in software development.&  \textbf{.7254}  &  .0447 \\ 
    I am better informed about the ethics and societal impacts of technology than most of my software developer peers.&  \textbf{.6803}  &  .0130 \\ 
    When working on computing projects with others, I could effectively voice my perspectives on ethical issues.&  \textbf{.8001}  &  .0194 \\ 
    I will promote an ethical practice in my workplace, even if initially met with resistance.&  .3056  &  .3361 \\ 
    There are processes within workplaces to handle reported ethical computing violations or concerns.&  .0571  &  \textbf{.6281} \\ 
    When I talk about ethical computing issues, my peers usually pay attention.&  .1395  &  \textbf{.5757} \\ 
    Software development professionals are allowed to have a say about ethical computing concerns at their workplaces.& -.0064  &  \textbf{.6206} \\ 
    When ethical computing concerns are raised by employees, workplaces are responsive to addressing these concerns.& -.0987  &  \textbf{.7679} \\ 
    \bottomrule
    \end{tabular}
    \caption{Exploratory Factor Analysis for the Critical Agency Subscales. Bolded factor loadings indicate loadings above the recommended minimum of .4.}
    \label{tab:EFA_critical_agency}
\end{table*}

\clearpage

\section{Confirmatory Factor Analysis Sample Statistics}
\label{appendix:sample_stats}

\begin{table*}[h!]
    \centering
    \begin{tabular}{lcc}
    \toprule
         & \textbf{Mean} & \textbf{Std. dev.} \\
    \midrule
    Independent Measure: \textit{Received ethics training}  & .62 &  \\
    \midrule
    Dependent Measures: Critical Reflection and Agency in Computing & & \\
    \textit{Valuing Ethics Training} & 4.63 & .98\\
    \textit{Valuing Technical Training} & 5.20 & .74 \\
    \textit{Techno-solutionism} & 3.63 & 1.12\\
    \textit{Valuing Marginalized Perspectives} & 4.36 & 1.20\\
    \textit{Personal Effectiveness} & 4.30 & 1.08\\
    \textit{System Responsiveness} & 4.07 & 1.04\\
    \midrule
    Demographic and Control Measures && \\
    \textit{Student status} & .58 & .49 \\
    \textit{Ethnicity} && \\
    \hspace{2mm} White & .48 & \\
    \hspace{2mm} Black & .20 & \\
    \hspace{2mm} Asian & .15 & \\
    \hspace{2mm} Mixed & .12 & \\
    \hspace{2mm} Other & .05 & \\
    \hspace{2mm} Prefer not to say & <.01 & \\
    \textit{Age} & 29.12 & 9.89 \\
    \textit{Gender} && \\
    \hspace{2mm} Man & .64 & \\
    \hspace{2mm} Woman & .31 & \\
    \hspace{2mm} Non-binary & .04 & \\
    \hspace{2mm} Man, Woman & .01 & \\
    \hspace{2mm} Woman, Non-binary & <.01 & \\
    \hspace{2mm} Genderqueer & <.01 & \\
    \bottomrule
    \end{tabular}
    \caption{Means and standard deviations of independent variables, dependent variables, and demographic and control measures (N = 474).}
    \label{tab:means_std_cfa}
\end{table*}

\clearpage

\section{Confirmatory Factor Analysis Loadings}
\label{appendix:cfa_loadings}

\begin{table*}[h!]
    \centering
    \begin{tabular}{p{9.5cm}ccc}
    \toprule
       \textbf{Latent Variable and Items} & \textbf{\makecell{Standardized \\ Estimate}} & \textbf{Std. Err.} & \textbf{$R^2$} \\
    \midrule
        \textbf{Factor 1}: ``Valuing Marginalized Perspectives'' ($\alpha$ = .63) & & & \\
        Considering issues of social justice should be a fundamental consideration in the design and development of any computing system. & .70 & .04 & .60 \\
        Developing computer software for public use requires input from marginalized groups.  & .67 & .04 & .47 \smallskip \\
        \textbf{Factor 2}: ``Techno-Solutionism'' ($\alpha$ = .79) & & & \\
        With enough resources, computing technologies can solve any problem.  & .73 & .03 & .44 \\
        Datasets that are large enough can overcome any bias in collection.   & .74 & .03 & .61 \\
        Biases in datasets can always be corrected with the right techniques.  & .69 & .04 & .39 \\
        Computing technologies benefit everyone equally. & .65 & .04 & .41 \\
    \bottomrule
    \end{tabular}
    \caption{Confirmatory Factor Analysis for the Critical Reflection Subscale: Recognizing computing/data embeds values/power.}
    \label{tab:CFA_computing_power}
\end{table*}

\begin{table*}[h!]
    \centering
    \begin{tabular}{p{9.5cm}ccc}
    \toprule
        \textbf{Latent Variable and Items} & \textbf{\makecell{Standardized \\ Estimate}} & \textbf{Std. Err.} & \textbf{$R^2$} \\
    \midrule
     \textbf{Factor 3}: ``Valuing Ethics Training'' ($\alpha$ = .87) & & & \\
     The social impacts of software. & .83 & .02 & .56\\
     The environmental impacts of software. & .70 & .03 & .51\\
     Legal considerations in software development. & .62 & .03 & .40 \\
     Ethical implications of topics being studied. & .82 & .02 & .63 \\
     Collaborating on software development projects with local community groups. & .65 & .03 & .36 \\
     Guidelines for discussing ethical issues with others. & .82 & .02 & .61 \\
     A software development code of ethics. & .76 & .02 & .44 \smallskip \\
     \textbf{Factor 4}: ``Valuing Technical Training'' ($\alpha$ = .87) & & & \\
     Computer architectures. & .66 & .03 & .43  \\
     Databases. & .72 & .03 & .52  \\
     Technical programming skills. & .71 & .03 & .50  \\
     Software quality assurance and testing. & .69 & .03 & .47  \\
     Computer science theory and algorithms. & .63 & .03 & .39  \\
     Identifying requirements to build software. & .72 & .03 & .51 \\
     Data structures. & .80 & .02 & .64  \\
        \bottomrule
    \end{tabular}
    \caption{Confirmatory Factor Analysis for the Critical Reflection Subscale: Recognizing computing training should include more explicit ethics and social impact discussions.}
    \label{tab:CFA_critical_reflection2}
\end{table*}

\begin{table*}[h!]
    \centering
    \begin{tabular}{p{9.5cm}ccc}
    \toprule
    \textbf{Latent Variable and Items} & \textbf{\makecell{Standardized \\ Estimate}} & \textbf{Std. Err.} & \textbf{$R^2$} \\
    \midrule
    \textbf{Factor 5}: ``Personal Effectiveness'' ($\alpha$ = .87) & & & \\
    I have a good understanding of the important ethical and social impacts to consider when developing software.& .78 & .02 & .61 \\
    I am able to participate in discussions about ethics and social impacts of computing.& .77 & .02 & .59 \\
    I am confident in my own ability to uphold ethical conduct in software development.& .76 & .02 & .58 \\
    I am better informed about the ethics and societal impacts of technology than most of my software developer peers.& .70 & .03 & .48 \\
    When working on computing projects with others, I could effectively voice my perspectives on ethical issues.& .82 & .02 & .67\smallskip\\
    \textbf{Factor 6}: ``System Responsiveness'' ($\alpha$ = .77) & & & \\
    There are processes within workplaces to handle reported ethical computing violations or concerns.& .72 & .03 & .51\\
    When I talk about ethical computing issues, my peers usually pay attention.& .67 & .03 & .45\\
    Software development professionals are allowed to have a say about ethical computing concerns at their workplaces.& .58 & .04 & .33\\
    When ethical computing concerns are raised by employees, workplaces are responsive to addressing these concerns.& .73 & .03 &  .53\\
    \bottomrule
    \end{tabular}
    \caption{Confirmatory Factor Analysis for the Critical Agency Subscales.}
    \label{tab:CFA_critical_agency}
\end{table*}

\clearpage

\end{document}